\documentclass[aps,prd,preprintnumbers,nofootinbib,twocolumn]{revtex4}
\usepackage{amsmath}
\usepackage{graphicx}
\usepackage{dsfont}
\usepackage{subfigure}

\def\3nab{\tilde{\nabla}}

\def\nn{\nonumber}

\def\hsp5{\hspace{5mm}}

\def\case#1/#2{\textstyle\frac{#1}{#2}}

\def\be {\begin{equation}}
\def\ee {\end{equation}}
\def\ber {\begin{eqnarray}}
\def\eer {\end{eqnarray}}
\def\bea {\begin{eqnarray}}
\def\eea {\end{eqnarray}}

\def\bc {\begin{center}}
\def\ec {\end{center}}
\def\case#1/#2{\frac{#1}{#2}}

\def\etal\;{{\it et al.}}

\begin{document}
\title{A characteristic signature of fourth order gravity}
\author{Kishore N. Ananda}
\affiliation{ Department of Mathematics and Applied\ Mathematics,
University of Cape Town, South Africa.}
\author{Sante Carloni}
 \affiliation{ Department of Mathematics and Applied\ Mathematics,
University of Cape Town, South Africa.}
 \author{Peter K. S. Dunsby}
\affiliation{ Department of Mathematics and Applied\ Mathematics,
University of Cape Town, South Africa.}
\affiliation{\ South African
Astronomical Observatory, Observatory Cape Town, South Africa.}
\begin{abstract}
We present for the first time the complete matter power spectrum for $R^n$ gravity which has been derived from the {\it fourth order} scalar perturbation equations.
This leads to the discovery of a characteristic signature of fourth order gravity in the matter power spectrum, the details of which have not seen before in other studies
in this area and therefore provides a crucial test for fourth order gravity on cosmological scales.
\end{abstract}
\pacs{04.50.+h, 04.25.Nx } \maketitle
Ever since the {\it Concordance model} \cite{concordance} was proposed as the best fit to all available cosmological data sets, there have been many attempts to understand the nature of Dark Energy. However, despite enormous effort over the past few years, this problem remains one of the greatest puzzles in contemporary physics. One of the theoretical proposals that has received a considerable amount of attention recently, is that Dark Energy has a geometrical origin. This idea has been driven by the fact that modifications to General Relativity appear in the low energy limit of many fundamental schemes \cite{stringhe,birrell} and that these modifications lead naturally to cosmologies which admit a Dark Energy like era \cite{SalvSolo} without the introduction of any additional cosmological fields.  Most of the work on this idea has focused on {\it fourth order gravity},  in which the standard Hilbert-Einstein action is modified with terms that are at most of order four in the metric tensor. The features of fourth order gravity have been analyzed with different techniques \cite{OurDynSys} and all these studies suggest that these cosmologies can give rise to a phase of accelerated expansion, which is considered to be an important footprint of Dark Energy.

The work described above has largely focused on the dynamics of homogeneous cosmologies which have the standard Robertson-Walker geometry and are therefore also isotropic. Although these results have many of the desirable features that we are looking for, such as a matter dominated epoch and late-time acceleration,  there are still some key issues that  need to be addressed before one could claim to have a cosmological description which is able to compete with the standard $\Lambda$CDM cosmology. The calculation and analysis of  the evolution of linear perturbations and their comparison with observations is clearly among the most important of these open problems. Over the past year this problem has been studied by several authors using the metric approach to perturbations,  by either considering different ways of parameterizing the non-Einstein modifications of gravity or by simplifying the underlying fourth-order perturbation equations using a quasi-static approximation \cite{Li:2008ai}.

In what follows, we demonstrate that  considerable progress can be made to this problem by using the {\it 1+3 covariant approach} to cosmological perturbations \cite{Covariant}. Using a specific recasting of the field equations (based on the Ricci and Bianchi identities),  the development of equations describing cosmological perturbations in  theories of gravity characterized by an action which is a general analytic function of the Ricci scalar $f(R)$, becomes both transparent and straightforward, allowing for the exact integration of the perturbation equations without making any approximations. In order to easily discuss the key features of the perturbation dynamics and the associated power spectrum, we focus on $R^{n}$-gravity. This theory is characterized by the action $L=\sqrt{-g}\left[\chi R^{n}+{\cal L}_{M}\right]$ and is the simplest possible example of fourth order gravity.

Before we can discuss the evolution of density perturbations, a suitable background cosmology must first be found. In \cite{dynsys05,SanteGenDynSys}, the complete dynamics  of homogeneous and isotropic cosmologies were studied in detail using the dynamical system approach (see \cite{ellisbook} and references therein).  It was found that in $R^n$ gravity, it is possible to have a transient matter-dominated decelerated expansion phase, followed by a smooth transition to a Dark Energy like era which drives the cosmological acceleration. The first phase, characterized by the baratropic equation of state parameter $w$, has an expansion history determined by a scale-factor $a=t^{2n/3(1+w)}$ where we restrict ourselves to $n>0$ for this background as negative values of $n$ would represent a contracting model.
This solution provides exactly the setting during which structure formation can take place and is therefore an ideal background solution for our study of density perturbations.

Scalar perturbations, which describe density perturbations may be extracted from any first order tensor $T_{ab}$ orthogonal to $u^a$ by using a  {\em local} decomposition \cite{EBH}, so that repeated application of the operator $\3nab_{a}\equiv h^{b}_{a}\nabla_b$ on $T_{ab}$ extracts the scalar part of the perturbation variables. In this way we can define the following scalar quantities
\begin{widetext}
\begin{equation}
\Delta_{m}=\frac{S^2}{\mu_{m}}\3nab^2\mu_{m}\,,\qquad
Z=S^2\3nab^2\Theta\,,\qquad C=S^{4}\3nab^2\tilde{R}\,,\qquad{\cal R}=S^2\3nab^2 R\,,\qquad\Re=S^2\3nab^2 \dot{R}\;.
\end{equation}
\end{widetext}
where $\Delta^m_a$,  $Z$ respectively represent the fluctuations in the matter energy density $\mu_m$ and expansion $\Theta$,  and ${\mathcal R}$,  $\Re$   determine the fluctuations in the Ricci scalar $R$ and its momentum $\dot{R}$.
This set of variables completely characterizes the evolution of density perturbations.  Then, using eigenfunctions of the  spatial Laplace-Beltrami operator defined in \cite{Covariant}: $\3nab^{2}Q = -\frac{k^{2}}{S^{2}}Q$,
where $k=2\pi S/\lambda$ is the wave number and $\dot{Q}=0$,  we can expand every first order quantity in the above equations:
\begin{equation}\label{eq:developmentdelta}
X(t,\mathbf{x})=\sum X^{(k)}(t)\;Q^{(k)}(\mathbf{x})\;,
\end{equation}
where $\sum$ stands for both a summation over a discrete index or an
integration over a continuous one.
In this way, it is straightforward, although lengthy, to derive a pair of second order equations describing the $k^{th}$ mode for density perturbations in $f(R)$ gravity. They are:
\begin{widetext}
\begin{eqnarray}
  &&\nn\ddot{\Delta}_{m}^{(k)}+\left[\left(
  \frac{2}{3}-w\right) \Theta -\frac{\dot{R}
  f''}{f'}\right] \dot{\Delta}_{m}^{(k)}-\left[w  \frac{k^2}{S^{2}}-w  (3
  p^{R}+\mu^{R})-\frac{2 w  \dot{R} \Theta
  f''}{f'}-\frac{\left(3 w ^2-1\right) \mu^{m} }{f'}\right]\Delta_{m}^{(k)}=\\&&=
  \frac{1}{2}(w +1)\left[-2 \frac{k^2}{S^2}\frac{f''}{f'}-1+
  \left(f-2 \mu^{m} +2 \dot{R} \Theta  f''\right)\frac{f''}{f'^2}
  -2  \dot{R} \Theta
  \frac{f^{(3)}}{f'}\right] \mathcal{R}^{(k)} -\frac{(w +1) \Theta
  f'' }{f'}\dot{\mathcal{R}}^{(k)}\label{EqPerIIOrd1}\,,\nonumber\\&&
  \nn f''\ddot{\mathcal{R}}^{(k)}+\left(\Theta f'' +2 \dot{R}
  f^{(3)}\right)
  \dot{\mathcal{R}}^{(k)}-\left[\frac{k^2}{S^2}f''+ 2 \frac{K}{S^2}f''
 +\frac{2}{9} \Theta^2 f''- (w +1) \frac{\mu^{m}}{2 f'}f''- \frac{1}{6}(\mu^{R}+ 3
p^{R})f''+\right.\\&&\nn\left.-\frac{f'}{3}+ \frac{f}{6 f'}f'' +
\dot{R} \Theta  \frac{f''^{2}}{ f'} -
  \ddot{R} f^{(3)}- \Theta f^{(3)} \dot{R}- f^{(4)}\dot{R}^2
  \right]\mathcal{R}^{(k)}=-
  \left[ \frac{1}{3}(3 w -1) \mu^{m}+ \right.\\&&\left.+\frac{w}{1+w} \left(f^{(3)}
  \dot{R}^2+  (p^{R}+\mu^{R}) f'+ \frac{7}{3}\dot{R} \Theta f''
+\ddot{R} f'' \right)\right]\Delta_{m}^{(k)}-\frac{(w -1) \dot{R} f''}{w +1} \dot{\Delta}_{m}^{(k)}\,.\label{EqPerIIOrd2}
\end{eqnarray}
\end{widetext}
where $f'=\partial f(R) / \partial R$, the quantities $\mu_R$, $p_R$ are the energy density and pressure of the {\it curvature fluid}  defined in \cite{SantePert}  and $K=0,+1,-1$ is the usual spatial curvature scalar . It is easy to see that for the $f(R)=R$ case, these equations reduce to the standard equations for the evolution of the scalar perturbations in General Relativity.

Already on super-Hubble scales, $k/aH\ll 1$, a number of important features are found which allows one to differentiate (\ref{EqPerIIOrd2}) from their General Relativity counterparts \cite{SantePert}. Firstly, it is clear that the evolution of density perturbations is determined by a {\it fourth order} differential equation rather than a second order one. This implies that the evolution of the density fluctuations contains, in general, four modes rather that two and can give rise to a more complex evolution than the one of General Relativity (GR).  Secondly, the  perturbations are found to depend on the scale for any equation of state for standard matter (while in General Relativity the evolution of the dust perturbations are scale-invariant). This means that even for dust, the evolution of super-horizon and sub-horizon  perturbations are different. Thirdly, it is found that the  growth of large density fluctuations can occur also in backgrounds in which the expansion rate is increasing in time (see figure 1). This is in striking contrast with what one finds in General Relativity and would lead to a time-varying gravitational potential, putting tight constraints on the Integrated Sachs-Wolfe effect for these models.

\begin{figure}[htbp]
\includegraphics[scale=0.7]{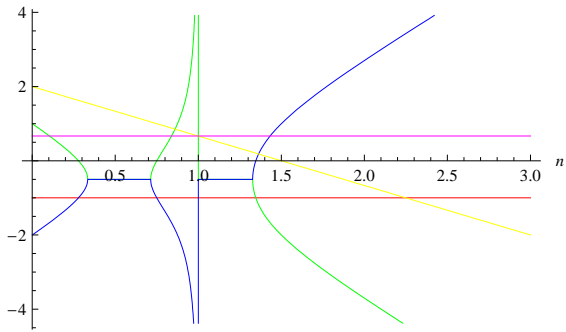}
\caption{Plot against $n$ of the real part of the long wavelength modes
for $R^n$-gravity in the dust case (blue, red green and yellow lines) together
with the GR modes (red and purple line). Note that there is at least one growing mode for any value of $n$. This means that even in cases where the expansion rate is accelerating, i.e., $n>\case{3}/{2(1+3)}$, large-scale density perturbations  grow.}
 \label{Fig1}
\end{figure}

Let us now turn to the case of a general wave mode $k$.  One of the most instructive way of understanding the details of the evolution of density perturbations for a general $k$ is to compute the matter power spectrum $P(k)$, defined by the relation \cite{Coles} $\langle\Delta_{m}({\mathbf k_1})\Delta_{m}({\mathbf k_2})\rangle=P(k_1) \delta({\mathbf k_1}+{\mathbf k_2})$, where ${\mathbf k_i}$ are two wavevectors characterizing two Fourier components of the solutions of (\ref{EqPerIIOrd1}) and $P({\mathbf k_1})=P(k_1)$ because of isotropy in the distribution of the perturbations. This quantity tells us how the fluctuations of matter depend on the wavenumber at a specific time and carries information about the amplitude of the perturbations (but not on their spatial structure). In General Relativity,  the power spectrum on large scales is constant, while on small scales it is suppressed in comparison with the large scales (i.e., modes which entered the horizon during the radiation era) \cite{PaddyPert}.  In the case of pure dust in General Relativity the power spectrum is scale invariant.  Substituting the details of the background, the values of the parameter $n$, the barotropic factor $w$, the spatial curvature index $K$ and the wavenumber $k$ into (\ref{EqPerIIOrd1}) one is able to obtain $P(k)$ numerically.

The k-structure of equations (\ref{EqPerIIOrd1})  suggest that in fourth order gravity there exist at least three different growth  regimes  of the perturbations. This is confirmed by our results (see figure 2). In particular, in the case of dust we have three regimes for any values of the remaining parameters:  (i) on very large scales the spectrum it is like what one finds for General Relativity, i.e., scale invariant;  (ii) as $k$ becomes bigger the scale invariance is broken and oscillations in the spectrum appear;  (iii)   for even larger $k$ the spectrum becomes again scale invariant. However, on these scales the spectrum can contain either an excess or deficit of power depending on the value of  $n$. In particular for $n\approx 1^+$ small scales have more power than large scales, but, as one moves towards larger values of $n$, the small scale modes are suppressed.
\begin{figure}[htbp]\label{PkRn1}
\includegraphics[scale=0.4]{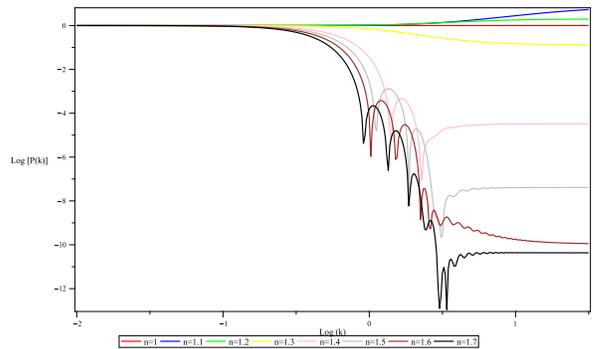}\label{Fig11}
\caption{Plot of  the power spectrum at $\tau=1$ for $R^n$-gravity and $n>1$. Note that the spectrum is composed of three parts corresponding to three different evolution regimes for the perturbations.}
\end{figure}

\begin{figure}[htbp]
\begin{center}
\includegraphics[scale=0.35]
{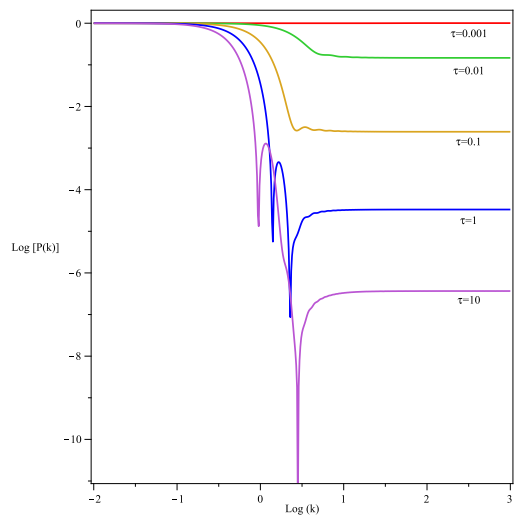}
\caption{Evolution the Power spectrum for $R^n$-gravity for $n=1.4$. The spectrum has been normalized in such a way  that the curves coincide at large scales. Note how, as time passes, small scale perturbations are dissipated and oscillation appear. \label{Fig15}}
\end{center}
\end{figure}

The features of the spectrum that we have presented above can be best interpreted by comparing the system (\ref{EqPerIIOrd2}) which produced it with the equations for the evolution of scalar perturbations for two interacting fluids in General Relativity \cite{DBE}. Immediately one notices that they have the same structure, i.e., there are friction and source terms due to the interaction and  the gravitation of the two effective fluids. It is then natural to ask ourselves if this analogy can be useful to better understand the physics of these models. The answer is affirmative. First of all a more correct way to draw this analogy would be to write the system of equations for  $\Delta_m$ and fluctuations in the energy density of the curvature fluid  $\Delta_R=S^{2}\3nab^2\mu_R/\mu_R$ and analyze their structure rather than using the ones above.  On very large and on very small scales, the coefficients of the  $(\Delta_m,\Delta_R)$ system become independent of $k$, so that the evolution of the perturbations does not change as a function of scale and the power spectrum is consequently scale invariant. On intermediate scales the interaction between the two fluids  is maximized and the curvature fluid acts as a relativistic component whose pressure is responsible for the oscillations and the dissipation of the small scale perturbations in the same way in which the photons operate in a baryon-photon system. This suggests the following interesting interpretation for the perturbation variables $\mathcal{R}$ and $\Re$. These quantities can be interpreted as representing the modes associated with the  contribution of the additional scalar degree of freedom typical of $f(R)$-gravity. In this sense the spectrum can be explained physically as a consequence of the interaction between  these scalar modes and standard matter. The result is a considerable loss of power for a relatively small variation of the parameter $n$.  For example, in the case $n=1.4$ the difference in power between the two scale invariant parts of the spectrum for $n=1.1$ is of one order of magnitude while for $n=1.6$  is about ten orders of magnitude. It should be noted that existing analysis\cite{CCT} of this model require $n=3.5$ in order for predictions to be consistent with measurements of rotation curves of low surface brightness galaxies and SNe Ia. Given the huge drop in power at small scales in the power spectrum for $n=3.5$ one expects that this model could be easily ruled out.

Further information on the dynamics of the matter perturbations can be obtained examining the time evolution of the power spectrum. In figure  \ref{Fig15} we give the power spectrum for $n=1.4$ at different times. One can see that, as the universe expands, the small scale part of the spectrum is more and more suppressed and oscillations start to form, suggesting that in this model small scale perturbations tend to be dissipated in time. On the other hand on large scales they do not evolve, which might appear in contrast with what mentioned above. However this is a byproduct of the normalization:  for clarity we have normalized the spectrum in such a way that every curve has the same power in long wavelength limit. A more in depth discussion of the features presented above can be found in \cite{Ananda}.

Probably the most important consequence of the form of the spectrum presented above is the fact that the effect of these type of fourth order corrections is evident only for a special range of scales, while the rest of the spectrum has the same $k$ dependence of GR (but different amplitude). This implies that we have a spectrum that both satisfies the requirement for scale invariance and has distinct features that one could in principle detect, by combining future Cosmic Microwave Background (CMB) and large scale surveys (LSS) \cite{Planck, SDSS}.

{\noindent{\bf Acknowledgements:}\\
The authors wish to thank Dr J. Larena for useful discussion and suggestions. KNA and SC are supported by Claude Leon Foundation fellowships. This work was supported by the National Research Foundation (South
Africa) and the {\it Ministrero degli Affari Esteri - DIG per la
Promozione e Cooperazione Culturale} (Italy) under the joint
Italy/South Africa science and technology agreement.


\begin{thebibliography}{99}
\bibitem{concordance}
J. P. Ostriker and P. J. Steinhardt, Cosmic Concordance,  [arXiv:astro-ph/9505066].
\bibitem{stringhe}D.~G.~Boulware and S.~Deser,
  Phys.\ Rev.\ Lett.\  {\bf 55}, 2656 (1985); J.~Z.~Simon,
  Phys.\ Rev.\  D {\bf 41} (1990) 3720; K.~Forger, B.~A.~Ovrut, S.~J.~Theisen and D.~Waldram,
  Phys.\ Lett.\  B {\bf 388}, 512 (1996)
  [arXiv:hep-th/9605145]; G.~Cognola, E.~Elizalde, S.~Nojiri, S.~Odintsov and S.~Zerbini,
  Phys.\ Rev.\  D {\bf 75}, 086002 (2007)
  [arXiv:hep-th/0611198].

\bibitem{birrell} N.D. Birrell and P.C.W. Davies, {\it Quantum Fields
in Curved Space}, Cambridge Univ. Press, Cambridge (1982).

\bibitem{SalvSolo}
  S.~Capozziello,
  Int.\ J.\ Mod.\ Phys.\  D {\bf 11}, 483 (2002)
  [arXiv:gr-qc/0201033]; S.~Capozziello, S.~Carloni and A.~Troisi,
 ``Recent Research Developments in Astronomy \&  Astrophysics"-RSP/AA/21 (2003)
  [arXiv:astro-ph/0303041].
\bibitem{OurDynSys}
  S.~Carloni, P.~K.~S.~Dunsby, S.~Capozziello and A.~Troisi,
  ``Cosmological dynamics of $R^n$ gravity'',
  Class.\ Quant.\ Grav.\  {\bf 22}, 4839 (2005)
  [arXiv:gr-qc/0410046]; T.~Clifton and J.~D.~Barrow,
  Phys.\ Rev.\  D {\bf 72}, 103005 (2005)
  [arXiv:gr-qc/0509059];
\bibitem{Li:2008ai}
B.~Li, J.~D.~Barrow, D.~F.~Mota and H.~Zhao,
arXiv:0805.4400 [gr-qc]; Y.~S.~Song, W.~Hu and I.~Sawicki,
Phys.\ Rev.\  D {\bf 75}, 044004 (2007)
[arXiv:astro-ph/0610532]; E.~Bertschinger and P.~Zukin,
  Phys.\ Rev.\  D {\bf 78}, 024015 (2008)
  [arXiv:0801.2431 [astro-ph]]; W.~Hu and I.~Sawicki,
Phys.\ Rev.\  D {\bf 76}, 104043 (2007)
[arXiv:0708.1190 [astro-ph]];
H.~Oyaizu, M.~Lima and W.~Hu,
 arXiv:0807.2462 [astro-ph].
\bibitem{Covariant}
G.~F.~R.~Ellis \& M.~Bruni
Phys Rev D {\bf 40} 1804 (1989);  M.~Bruni,  P.~K.~S.~Dunsby \& G.~F.~R.~Ellis,
Ap. J. {\bf 395} 34 (1992).
\bibitem{dynsys05} S. Carloni, P. Dunsby, S. Capozziello \& A. Troisi
{\it Class. Quant. Grav.} {\bf 22}, 4839 (2005).
\bibitem{SanteGenDynSys}  S. Carloni, A. Troisi and P. K. S. Dunsby,
[arXiv:0706.0452].  To appear in General Relativity and Gravitation (2009).

 \bibitem{ellisbook} {\em Dynamical System in Cosmology} edited by
Wainwright J and  Ellis G F R (Cambridge: Cambridge  Univ. Press
1997) and references therein.
\bibitem{EBH}
G.~F.~R.~Ellis, M.~Bruni and J.~Hwang,
 Phys.\ Rev.\  D {\bf 42} (1990) 1035 (1990).
 \bibitem{SantePert}
  S.~Carloni, P.~K.~S.~Dunsby and A.~Troisi,
  ``The evolution of density perturbations in $f(R)$ gravity,'' Phys. Rev. D 77, 024024 (2008)
  arXiv:0707.0106 [gr-qc].
  \bibitem{Coles}
  P.~Coles and F.~Lucchin,
{\it  Chichester, UK: Wiley (1995) 449 p}

  \bibitem{PaddyPert} T.~Padmanabhan,
  AIP Conf.\ Proc.\  {\bf 843} (2006) 111
  [arXiv:astro-ph/0602117]; T.~Padmanabhan ``Structure Formation in the Universe" Cambridge university press (Cambridge)

  \bibitem{DBE}
   P.~K.~S.~Dunsby, M.~Bruni and G.~F.~R.~Ellis,
  Astrophys.\ J.\  {\bf 395}, 54 (1992).

\bibitem{CCT}
  S.~Capozziello, V.~F.~Cardone and A.~Troisi,
  Mon.\ Not.\ Roy.\ Astron.\ Soc.\  {\bf 375} (2007) 1423
  [arXiv:astro-ph/0603522].

\bibitem{Ananda}
  K.~N.~Ananda, S.~Carloni and P.~K.~S.~Dunsby,
  arXiv:0809.3673 [astro-ph].

   \bibitem{Planck}
    [Planck Collaboration],
  arXiv:astro-ph/0604069.

\bibitem{SDSS} see the webpage http://www.sdss.org/

\end{thebibliography}
\end{document}